\documentclass[12pt,a4paper]{article} 
\usepackage{epsfig}
\usepackage{graphics}
\usepackage{a4wide}
\usepackage{amsmath}

\title{Stabilizing Textures with Magnetic Fields.}
 \author{R S Ward\footnote{email: richard.ward@durham.ac.uk}
 \bigskip
\\Department of Mathematical Sciences,  \\ University of
Durham, \\Durham DH1 3LE}

\newcommand{\cf}{{\it cf.}}
\newcommand{\half}{{\scriptstyle\frac{1}{2}}}
\newcommand{\quar}{{\scriptstyle\frac{1}{4}}}
\newcommand{\eighth}{{\scriptstyle\frac{1}{8}}}
\newcommand{\RR}{{\bf R}}
\newcommand{\CC}{{\bf C}}
\newcommand{\cE}{{\cal E}}
\newcommand{\cL}{{\cal L}}
\newcommand{\vecp}{\vec\phi}
\newcommand{\pa}{\partial}
\newcommand{\ii}{{\rm i}}
\renewcommand{\a}{\alpha}
\renewcommand{\b}{\beta}
\newcommand{\G}{\Gamma}
\renewcommand{\l}{\lambda}
\newcommand{\s}{\sigma}

\begin{document}
\maketitle \abstract{\noindent
The best-known way of stabilizing textures is by Skyrme-like terms, but
another possibility is to use gauge fields.  The semilocal vortex may be
viewed as an example of this, in two spatial dimensions.  In three dimensions,
however, the idea (in its simplest form) does not work --- the link
between the gauge field and the scalar field is not strong enough to prevent
the texture from collapsing.  Modifying the $|D\Phi|^2$ term in the
Lagrangian (essentially by changing the metric on the $\Phi$-space) can
strengthen this link, and lead to stability.  Furthermore, there is a limit
in which the gauge field is entirely determined in terms of the scalar field,
and the system reduces to a pure Skyrme-like one.  This is described for
gauge group U(1), in dimensions two and three.  The non-abelian version is
discussed briefly, but as yet no examples of texture stabilization are known in
this case.
}
\newpage

\section{Introduction}

Textures are classical solutions which are characterized
by a nonzero homotopy group $\pi_d(T)$, $d$ being the number of space
dimensions.  The relevant systems typically involve a scalar field
$\Phi$ taking values in the target space $T$.
With a Lagrangian such as $|\pa_\mu\Phi|^2$, and for
$d\geq2$, configurations are prone to implode (by the usual
Derrick scaling argument).  In an expanding universe, textures might be
stabilized by the cosmological expansion; but we are interested here
in cases where gravitational effects are negligible, and we take space-time
to be flat.  In flat space, the best-known way of stabilizing textures
is to add a Skyrme term involving higher powers of $\pa_\mu\Phi$.

By contrast, vortices or monopoles correspond to a nontrivial
$\pi_{d-1}(T)$, and (in their ``local'' versions) are stabilized
by gauge fields.  Many similarities between textures and
vortices/monopoles have been noted.
For example, multi-Skyrmions and BPS multi-monopoles (located at a single
point in space) each have a polyhedral structure corresponding to an
appropriate subgroup of O($d$), and this has been partly understood in
terms of rational maps from the Riemann sphere to itself \cite{HMS98}.
The purpose of this paper is to investigate the stabilization of
textures by gauge fields, and so in particular it explores a 
different sort of relation between the two classes of
topological solitons, generalizing the example provided by the
semilocal vortex \cite{VA91,P92}.

The idea of stabilizing textures with gauge fields has been
investigated before.  One motivation
has been the fact that Skyrme terms are non-renormalizable,
whereas gauge theories may have better quantum behaviour;
but in this paper the considerations are entirely classical.
For the extended abelian-Higgs model (with the Higgs field being a complex
doublet), it was pointed out in \cite{H93}
that an expansion in field gradients produced a Skyrme-like term, which
suggested stability; at the time, this was not investigated in detail.
More recent numerical simulations \cite{NPV00} seemed to show that
stability was indeed present (although, as reported below, we have not been
able to confirm this result).  In a different abelian system (involving
a triplet of real scalar fields and a massive abelian gauge field) no stable
textures could be found \cite{PT00}.  For the non-abelian case, scaling
arguments again suggest stability (\cf\ \cite{P00}); but detailed
investigation such as \cite{BTT96} have produced negative results.
The conclusion seems to be that the scalar field and the gauge field
have to be linked to each other sufficiently strongly in order to prevent
each from collapsing independently; and in ``standard'' systems, this link is
not strong enough.

The general framework is as follows.  Suppose we have a system involving
a gauge field (with gauge group $G$), and a multiplet $\Phi$ of scalar
fields coupled to it.  The ``basic'' Lagrangian of the system has the form
\begin{equation} \label{basic}
 \cL=\half\left|D_\mu\Phi\right|^2 -\quar\left(F_{\mu\nu}\right)^2-V(\Phi)\,.
\end{equation}
For space dimension $d=2$, the system defined by (\ref{basic}) may admit
stable static solutions (for example, semilocal vortices); but for $d=3$
it seems not to --- some modification is needed.  The idea pursued here
is that the term $|D_\mu\Phi|^2$ in the Lagrangian involves a choice of
metric on the space $T$ in which $\Phi$ takes its values, and we can change
this metric.  For example, if $\Phi$ is a complex vector, then the standard
Euclidean metric is $|D\Phi|^2 = (D\Phi^{\dagger})(D\Phi)$, where
$D\Phi^{\dagger}$ denotes the complex-conjugate transpose of $D\Phi$.
A natural modification of this (see the following section) is to add a term
$\kappa^2|\Phi^{\dagger}D_\mu\Phi|^2$, where $\kappa$ is a constant. So we now
have a family of systems, parametrized by $\kappa$.  Taking the limit
$\kappa\to\infty$ enforces the constraint
\begin{equation} \label{constraint}
 \Phi^{\dagger}D_\mu\Phi=0\,,
\end{equation}
which (under favourable circumstances) determines the gauge potential
in terms of~$\Phi$.  So we have a  family of systems where, in an
appropriate limit, the gauge degrees of freedom disappear, and the
Maxwell/Yang-Mills term $(F_{\mu\nu})^2$ becomes a Skyrme term.  This
enables us to track a soliton solution as it changes from
a gauge-stabilized texture into a Skyrme-stabilized texture.
In an appropriate limit of parameters $\kappa,\ldots\to\infty$,
one gets a Skyrme system which certainly admits stable solitons; one
question is for which {\em finite} values of these parameters
there are stable solitons.

Non-trivial examples of this idea have only been found in the abelian
case $G={\rm U(1)}$, and these are described in the next two sections
(for $d=2$ and $d=3$ respectively).  A discussion of the non-abelian
case $(G={\rm SU(2)}$ in $d=3$) is given in the final section.
The conclusion therefore is that textures can be stabilized by (abelian)
magnetic fields, but no non-abelian version of this appears to be known.

It might be noted that the idea of adding a term
$\kappa^2|\Phi^{\dagger}D_\mu\Phi|^2$, and investigating how solitons depend
on the parameter $\kappa$, has been investigated before; the simplest
example (in a somewhat different context)
is that of the $CP^1$ model with no gauge field \cite{Z86}.


\section{Semilocal Vortices and Planar Skyrmions.}

In this section, we take $d=2$ (so space is the plane $\RR^2$), and
gauge group U(1).  Let the Higgs field $\Phi$ be a complex doublet
$\Phi = [\Phi^1\quad \Phi^2]^t$.  The resulting extended abelian-Higgs
system admits semilocal vortex solutions \cite{VA91,P92}; and in the limit
$\kappa\to\infty$ it becomes, as we shall see, a Skyrme version of the
$CP^1$ model.  The generalization with $\Phi$ being an $M$-tuplet, leading
in the limit to a Skyrme version of the $CP^{M-1}$ model, is
straightforward; but for simplicity we shall restrict here to the $CP^1$
case.

The standard Lagrangian is
\begin{equation} \label{SLV-Lag}
 \cL = \half\left(D_\mu\Phi\right)^{\dagger}\left(D^\mu\Phi\right)
       - \quar\left(F_{\mu\nu}\right)^2
       - \eighth\l(1 - \Phi^{\dagger}\Phi)^2\,,
\end{equation}
where $D_\mu\Phi = \pa_\mu\Phi - \ii A_\mu\Phi$.  For the semilocal vortex
solution, the gauge field provides a ``hard core''
which prevents the soliton from shrinking.  If $0<\l<1$, the
single soliton is stable; but for $\l>1$ it is unstable (it expands
without limit) \cite{H92,AKPV92}.  For $\l=1$ there is a one-parameter
family of static solutions saturating a Bogomolny bound, but these solitons
are marginally unstable \cite{L92}.  One member of this family
is (an embedding of) the standard Nielsen-Olesen vortex.

Various relations between this system and the $CP^1$ model have
been noted before (\cf\ \cite{H93,S96}).
For example, imposing the constraint $\Phi^{\dagger}\Phi=1$ (this
corresponds to letting the parameter $\l$ tend to infinity), and scaling
away the $(F_{\mu\nu})^2$ term, leaves the $CP^1$ model \cite{S96}.
But in order to have stable semilocal vortices which become Skyrmions as a
limiting case, one needs to make some modifications.

Recall, first, the symmetry of this system \cite{P92}.  The ungauged system
has an SO(4) global symmetry.  On gauging a U(1) subgroup, this SO(4) is
reduced to the product of the local U(1) and a global SU(2); the field
$\Phi$ belongs to the fundamental representation of this SU(2).
The most general SU(2)-invariant metric on $T=\CC^2$ is
$h_{PQ}\,D\Phi^P\,D\bar\Phi^Q$, where
\begin{equation} \label{metric}
 h_{PQ}(\Phi,\bar\Phi) = g(\xi)\,\delta_{PQ} + \tilde g(\xi)\,\bar\Phi_P\Phi_Q
\end{equation}
and $\xi = \bar\Phi^P\Phi^P = \Phi^{\dagger}\Phi$.  The two functions $g$
and $\tilde g$ are arbitrary.  But in the limit $\l\to\infty$, which is of
particular interest here, we have $\xi\equiv1$; so let us take $g$ and
$\tilde g$ to be constants, scaling $g$ to unity and writing
$\tilde g=\kappa^2$.  Using this modified metric instead
of the standard Euclidean one amounts to replacing (\ref{SLV-Lag}) by
\begin{equation} \label{SLV-Lag2}
 \cL = \half\left(D_\mu\Phi\right)^{\dagger}\left(D^\mu\Phi\right)
       + \half\kappa^2|\Phi^{\dagger}D_\mu\Phi|^2
       - \quar\left(F_{\mu\nu}\right)^2
       - \eighth\l(1 - \Phi^{\dagger}\Phi)^2\,.
\end{equation}

The second modification is as follows.  In order to have stability for
$\l>1$, we need an extra potential term, which necessarily breaks the SU(2)
global symmetry (see for example \cite{AP97,APT98}).  We shall add to
(\ref{SLV-Lag2}) the term $\a|\Phi^2|^2$, where $\a$ is a positive constant.
In the Bogomolny case ($\kappa=0$ and $\l=1$), there is now a unique minimium:
it has $\Phi^2=0$, and is the Nielsen-Olesen vortex with energy $E=\pi$.

With these two modifications, the static energy density $\cE$ of the system is
given by
\begin{equation} \label{SLV-En}
 2\cE = \left(D_j\Phi\right)^{\dagger}\left(D_j\Phi\right)
     + \kappa^2|\Phi^{\dagger}D_j\Phi|^2 + \left(B_j\right)^2 + V(\Phi)\,,
\end{equation}
where $V(\Phi)= \quar\l(1-\Phi^{\dagger}\Phi)^2+2\a|\Phi^2|^2$, and
where $B_j=\epsilon_{jkl} \pa_k A_l$ is the magnetic field strength.

The boundary conditions are chosen to ensure finite energy.
At spatial infinity, one must have
\begin{description}
\item[(a)] $A_j = f^{-1}\pa_j f$, where $|f|=1$;
\item[(b)] $D_j\Phi = 0 \Rightarrow \Phi = f^{-1}K$,
          where $K$ is a constant 2-vector;
\item[(c)] $V(\Phi)=0 \Rightarrow K=[k\quad 0]^t$ with $|k|=1$.
\end{description}
Because of (b) and (c), $\Phi$ cannot be zero at spatial infinity; and
in order for $\Phi$ to be single-valued, $f$ has to be single-valued.
Hence $f$ is a map from the circle at spatial infinity to the gauge group
U(1), and the degree of $f$ is the soliton number~$N$.  The total magnetic
flux is proportional to~$N$, in the usual way.
The fact that there is nontrivial topology does not necessarily mean that
there are stable solitons; but the numerical work described below indicates
that there are, at least for certain ranges of the parameters $\a$,
$\l$ and $\kappa$.

Taking the limit $\l\to\infty$ enforces the constraint
$\Phi^{\dagger}\Phi=1$ (so $\Phi$ takes values in $S^3$).  If in addition
$\kappa\to\infty$, then the minimum-energy configuration approaches one
for which $\Phi^{\dagger}D_j\Phi=0$, and hence
\begin{equation} \label{gauge}
        A_j = -\ii \Phi^{\dagger}\pa_j\Phi.
\end{equation}
With $A_j$ is given in terms of $\Phi$ by this expression,
$(D_j\Phi)^{\dagger}(D_j\Phi)$ becomes the standard $CP^1$ energy,
and $(B_j)^2$ becomes a Skyrme term.  We can re-express this
as an O(3) sigma-model in the usual way: define a unit 3-vector
field $\vecp$ by $\vecp = \Phi^{\dagger}{\vec\sigma}\Phi$, where $\s^a$
are the Pauli matrices.  This corresponds to the standard Hopf map from
$S^3$ (the space $\Phi^{\dagger}\Phi=1$) to $S^2$. Strictly speaking,
the $\Phi$ field is
a vortex (winding at spatial infinity); but the $\vec\phi$ field, obtained
from it by projection, is a texture (constant at spatial infinity).
Then in the
$\l,\kappa\to\infty$ limit, the energy density $\cE$ is given by
\begin{equation} \label{OBS-En}
 8\cE_{\l,\kappa\to\infty} = (\pa_j\vec\phi)\cdot(\pa_j\vec\phi)
   + (\vec\phi\cdot\pa_1\vec\phi\times\pa_2\vec\phi)^2
   + 4\a(1-\vec n\cdot\vec\phi),
\end{equation}
where $\vec n$ is a constant unit vector.
This is a planar Skyrme system \cite{PSZ95a,PSZ95b}.
The energy of the Skyrmion solutions depends on $\a$, and can
be found by numerical minimization; for the 1-soliton with $\a=1$ it is
$E=3.1557\pi$.

The energy (and the stability) of the solitons in the system (\ref{SLV-En})
may be investigated numerically, as a function of the three parameters
$\a$, $\l$ and $\kappa$, and of the soliton number $N$.  This has been
done for the $N=1$ case, with $\a=1$ and $\l=1+\kappa^2$.  The result is
summarized in Figure 1, which shows the energy $E$
as a function of $\kappa\geq0$.  It was obtained by assuming
the standard form for O(2)-symmetric fields, namely
$\Phi^1=f(r)\exp(\ii N\theta)$, $\Phi^2=g(r)$, $A_r=0$ and
$A_{\theta}=a(r)$, where $f$, $g$ and $a$ are real-valued.  The
discrete version of the energy functional $E[f,g,a]$ was then minimized
numerically, using a conjugate-gradient method.  For each value of $\kappa$,
a stable solution was found.  Note that, as expected, $E$ goes from $E=\pi$
(the Nielsen-Olesen vortex) at $\kappa=0$ and $\l=1$, to $E=3.1557\pi$ (the
planar Skyrmion) as $\kappa\to\infty$ and $\l\to\infty$.
\begin{figure}[htb]
\begin{center}
\includegraphics[scale=0.4]{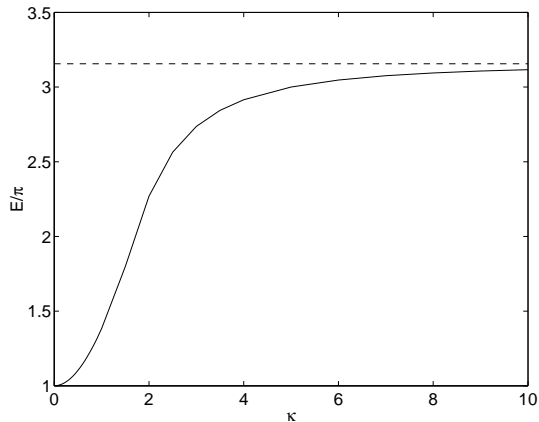}
\caption{The energy $E$ of the 1-soliton on $\RR^2$, as a function of
  $\kappa$, with $\a=1$ and $\l=1+\kappa^2$.  The dashed line is
  the energy of the planar Skyrmion obtained in the limit $\kappa\to\infty$.}
\label{fig1}
\end{center}
\end{figure}
%
%


\section{Vortex Rings and Hopf Textures.}

In this section, we investigate the same system (\ref{SLV-Lag2}) as before,
but in spatial dimension $d=3$.  The extra potential term
is omitted (in other words, $\a=0$), so the global SU(2) symmetry is
unbroken.  One may form a texture configuration by taking a finite length
of semilocal vortex with its ends joined together to form a loop in 3-space,
and it has previously been speculated that such a texture might be stable
\cite{H93,NPV00}.

The energy density is given by (\ref{SLV-En}), with $\a=0$; so the
system depends on the two parameters $\l$ and $\kappa$.  In the limit
$\l,\kappa\to\infty$, we again get an $S^3$-valued scalar field $\Phi$,
with the gauge potential being given by (\ref{gauge}); it has previously
been pointed out (\cf\ \cite{BW01,BFN02}) that this limit is equivalent
to the Faddeev-Hopf system \cite{FN97,GH97,BS99,HS99,HS00,W99,W00}.
So there are stable ring-like solitons in the limit; the question here is
whether they are stable for finite values of $\kappa$ and $\l$.

The boundary conditions imply, as before, that
$\Phi = [\Phi^1\quad\Phi^2]^t = f^{-1}K$ at $r=\infty$,
where $K$ is a constant 2-vector; so $W=\Phi^1/\Phi^2$ is constant
at spatial infinity.  Thinking of $W$ as a stereographic coordinate
for $CP^1$ therefore shows that $\Phi$ (provided it is nowhere-zero)
defines a map from $\RR^3$ to $S^2$ which is constant at infinity, and
hence is classified topologically by the Hopf number $N\in\pi_3(S^2)$.
For $N=1$, the field resembles a single vortex ring.
\begin{figure}[bht]
\begin{center}
\includegraphics[scale=0.4]{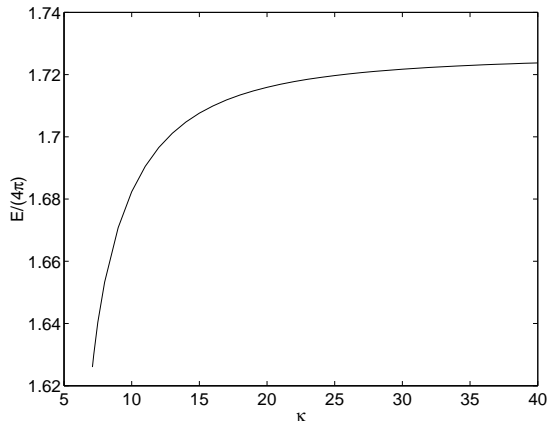}
\caption{The energy $E$ of the 1-soliton on $\RR^3$, as a function of
  $\kappa$, with $\l=1+\kappa^2$.}
\label{fig2}
\end{center}
\end{figure}
The stability of such $N=1$ configurations has been investigated numerically,
again by minimization of the energy functional.  The solitons cannot be
spherically-symmetric, but one expects that for small values of $N$
they will be axially-symmetric \cite{GH97,BS99}.  So one can reduce the
problem to a two-dimensional one which is not too difficult computationally.
More precisely, one can use cylindrical coordinates, and impose an
SO(2)$\times$SO(2)-invariant ansatz, as for example in \cite{NPV00}.

Minima were sought for the one-parameter family of systems obtained by
setting $\l=\kappa^2+1$, and stable solitons were found for $\kappa\geq7.1$.
Their energy is plotted in Figure 2.  For $\kappa\leq7$, however, the radius
of the vortex ring shrinks to zero, and the field unwinds: there is no stable
minimum.  When $\kappa$ (and therefore $\l$) tend to infinity, the normalized
energy $E'=E/4\pi$ approaches the value $E'_{\infty}=1.73$ (obtained by
extrapolation of the data in Figure 2).  This is exactly the energy of the
single Hopf soliton: from \cite{W00}, and allowing for different coupling
constants, we get the value $E'=1.22\sqrt{2}=1.73$.

The analogous computation previously reported in \cite{NPV00}
for the $\kappa=0$ case suggested that one might have
stability for fairly small values of $\l$ (of order unity).
The results described above do not confirm this, and in fact no stable
solution could be found for $\kappa=0$, even with $\l$ quite large.
But (as emphasized in \cite{NPV00}), there might be local minima
in the configuration space which are difficult to detect, and which
require an initial condition which is very close to the actual solution.
So it remains an open question as to whether stable vortex rings exist
for small values of $\kappa$ and $\l$.  It is, however, the case that
the configuration which is stable for $\kappa=7.1$, $\l=51.4$ collapses if 
$\kappa$ and $\l$ are reduced to $\kappa=7$, $\l=50$.


\section{Non-Abelian Gauge Field.}

As mentioned in the introduction, the question of whether textures can
be stabilized by a non-Abelian gauge field has previously been investigated;
there are suggestions based on simple scaling arguments (\cf\ \cite{P00}),
but more detailed studies have yielded negative results (\cf\ \cite{BTT96}).
Let us look at the three-dimensional case ($d=3$), with gauge group SO(3).
The field $\Phi$ belongs to some representation $\G$ of SO(3); so we have
to choose $\G$, as well as an appropriate potential function $V(\Phi)$.
For example, for the 'tHooft-Polyakov monopole one uses the fundamental
representation $\G = {\bf3}$.  The simplest extension of this is
the four-dimensional representation $\G = {\bf1}\oplus{\bf3}$.
The corresponding system admits monopole-like soliton solutions which
have been referred to as semilocal monopoles \cite{P92}. (Another simple
extension is $\G = {\bf3}\oplus{\bf3}$, the
corresponding solitons being referred to as coloured monopoles \cite{AKPV92}.)

Let us look at the ${\bf1}\oplus{\bf3}$ case: so $\Phi=(\phi_0,\vec\phi)$
is a four-vector.
Take the potential function to be $\l(1-|\Phi|^2)^2$; so for large
$\l$, we get the constraint $|\Phi|^2\approx1$.  One may then 
impose texture boundary conditions (rather than monopole
boundary conditions): namely, $\Phi$ tends to a constant as $r\to\infty$
in $\RR^3$.  So $\Phi$ is effectively a map from $S^3$ to $S^3$, and
it has a winding number $N$.  The stability of spherically-symmetric $N=1$
configurations has been studied numerically --- the details are as follows.

For simplicity, we shall take the $\l\to\infty$ limit, so
$|\Phi|^2\equiv1$; and the metric on $\Phi$-space to be flat (no extra
term analogous to $\kappa^2|\Phi^{\dagger}D_j\Phi|^2$).  The energy density is
\begin{equation}
\cE = \half|D_j\Phi|^2 + \quar(F_{jk})^2,
\end{equation}
where $D_j\Phi = (\pa_j\phi_0, \pa_j\phi^a-2\epsilon^{abc}A_j^b\phi^c)$.
To implement spherical symmetry, we take $\Phi$ and the gauge potential
$A$ to have the standard ``hedgehog'' form
\begin{equation}
 A^a_j=\epsilon_{jak}x^k f(r)/r^2, \quad \phi_0=\cos{g(r)},
      \quad \phi^a = x^a\sin{g(r)}/r\,,
\end{equation}
with the boundary conditions $f(0)=0$, $f(\infty)=\half$,
$g(0)=\pi$, $g(\infty)=0$.  The energy density then becomes
\begin{equation}
\cE = \frac{f_r^2}{r^2} + \frac{2f^2(f-1)^2}{r^4}
        + \frac{g_r^2}{2} + \frac{\sin^2{g}}{r^2}\left[1+4f(f-1)\right].
\end{equation}
One can then minimize the energy numerically; this was done using a
conjugate-gradient method, with various initial conditions.
But no smooth minimum could be found --- in every case, both $f$ and $g$
collapse towards being zero almost everywhere.

One can see this collapse analytically, in the following highly-simplified
version (involving just two degrees of freedom $\a$ and $\b$).
Let $\a$ and $\b$ be the values of $r$ such that $f(\b)=1/4$ and
$g(\a)=\pi/4$.  In other words, $\a$ and $\b$ are the ``radii'' of the
scalar field and the gauge field respectively.
More explicitly, take $f$ and $g$ to have the form
\[
f(r) = \left\{\begin{array}{ll}
              \frac{r^2}{4\b^2} & \mbox{for $0\leq r\leq\b$} \\
              \frac{1}{2}-\frac{\b}{4r} & \mbox{for $r\geq\b$}
              \end{array}
       \right\},
\quad
\cos{g(r)} = \left\{\begin{array}{ll}
                     -1+\frac{r^2}{\a^2} & \mbox{for $0\leq r\leq\a$} \\
                      1-\frac{\a^2}{r^2} & \mbox{for $r\geq\a$}
                    \end{array}
       \right\}.
\]
One can compute the energy $E(\a,\b)$ of this configuration exactly: it is
a rational function of $\a$ and $\b$.  In particular, for $\b=1/\a$ the
energy has the form $E(\a,1/\a) = \a\times\mbox{(polynomial in $\a$)}$.
The salient point about this form is that its minimum occurs
when $\a=0$; this corresponds to the scalar field shrinking to zero
width, while the gauge field spreads out.  As one sees from the usual
Derrick scaling argument used in \cite{P00}, the contribution to
the energy from the $|D_j\Phi|^2$ term can be reduced by scaling one way,
while the  contribution from $(F_{jk})^2$ can be reduced by scaling the
other way.  But the system as a whole can never reach a balance:
the gauge field and the scalar field are just not sufficiently strongly
coupled to each other to prevent each one from collapsing separately.
As was remarked before, these results do not actually prove the
absence of a stable solution: there might still be a local minimum
somewhere in the configuration space.  But it seems rather unlikely
that this system does admit a stable texture.

As in the previous sections, one can modify the metric on $\Phi$-space,
and this may improve the stability properties.  That possibility has not yet
been fully investigated; but certainly there is no gauge-invariant extra term,
the vanishing of which determines the gauge potential as in the abelian case
(\ref{gauge}).  So the idea of obtaining the usual Skyrme model as a limit
does not quite work in this non-abelian case.  Using other representations
$\G$, and for that
matter other gauge groups, opens up many more possibilities, which are still
to be explored.   But for the time being, it
remains the case that there are no known examples of three-dimensional
systems in which a texture is stabilized by a non-abelian gauge field.

\end{document}